\documentclass[conference]{IEEEtran}
\IEEEoverridecommandlockouts

\usepackage{cite}
\usepackage{amsmath,amssymb,amsfonts}
\usepackage{algorithmic}
\usepackage{graphicx}
\usepackage{xcolor}
\usepackage{textcomp}
\usepackage{colortbl}
\usepackage{fancyhdr}
\usepackage{booktabs}
\usepackage[keeplastbox]{flushend}
\usepackage{marvosym}
\usepackage{listings}
\usepackage{color} 
\definecolor{mygreen}{RGB}{28,172,0} 
\definecolor{mylilas}{RGB}{170,55,241}

\def\BibTeX{{\rm B\kern-.05em{\sc i\kern-.025em b}\kern-.08em
    T\kern-.1667em\lower.7ex\hbox{E}\kern-.125emX}}
\begin{document}

\title{Robust Control Approaches for Minimizing the Bandwidth Ratio in Multi-Loop Control\\
\thanks{I am thankful to Prof. Samuel Burden for his insightful discussions and suggestions. The idea of optimal bandwidth separation came up in one such discussion.}
}
\vspace{-1cm}
\author{\IEEEauthorblockN{Rahul Mallik}

\IEEEauthorblockA{Department of Electrical and Computer Engineering, University of Washington, Seattle, WA 98195, USA\\
Email: rmallik@uw.edu}
}

\IEEEoverridecommandlockouts
\IEEEpubid{\makebox[\columnwidth]{978-1-7281-1842-0/19/\$31.00~\copyright2019 IEEE \hfill} \hspace{\columnsep}\makebox[\columnwidth]{ }}
\maketitle
\IEEEpubidadjcol
\begin{abstract}
Conventional dc-dc, dc-ac converter control entail an inner current and outer voltage (ICOV) control loop.
Stability of multi-loop control is achieved by ensuring faster dynamics of inner loops, often separating the bandwidths by large factors. 
Heuristically a \emph{factor-of-ten} has often been used to separate the bandwidths of two adjacent loops in the nested architecture.
In this paper, we present a numerical method to optimally select the ratio of bandwidths for a nested controller.
Inspired from~\cite{Johnson2015OptimalSF}, we first manipulate the robust framework to show that the optimal $H_\infty$ subsumes the classical inner current outer voltage control.
We then use optimality of the proposed $H_\infty$ controller to inspect feasibility of different ratios of bandwidths of the inner and outer controllers.
We finally select the numerically closest bandwidths of the two nested loops which the guarantee stability. 
Simulation models are used to verify the optimal bandwidth separation for a candidate grid-forming dc-ac converter with inner current-outer voltage control.
\end{abstract}


\section{Introduction}\label{sec:Intro}

Power converters have become more flexible with nuanced devices like IGBTs, MOSFETs which can handle way more complex and fast switching transients than the traditional thyristors. This means, that if we now come up with some fancy controllers, then we are sure that the devices to act on them, are equally fast. The development of digital controllers is another boom in the field of power electronics. Even for a simple PI controller, one would have needed four op-amps, imagine what would be the need to develop a full blown 12 order digital controller (as a matter of fact, we do that in this project). SO simplicity was the key back then. With the advent of Field Programmable Gate Arrays (FPGAs), a lot has changed. One is no longer dependent on values of resistors and capacitors to tune a PI controller gain. We now have microcontrollers with 32 bit precision and we can specify and number of controllers parameters to the highest degree of precision. Classical control for SISO system draws heavily from loop shaping and similar designs which was more arts before Bode came and formalized with his theorems. The key principle was to maintain a flat high gain loop shape in the regions where we needed reference tracking or disturbance rejection and then roll it off at higher frequency where noise dominates.

To achieve this, small order controllers were just enough. But imagine if we want the loop shape o look really fancy, wiggly and we specify requirements at each frequency, and we do not want a generic high gain at other frequency because we are sure those frequencies would neither have reference signals or disturbances. This alone meant looking for tools where those specifications could be meaningfully entered and one which would synthesize an equally meaningful controller. It is not difficult to see why the focus has shifted from intuitive non rigorous design techniques of loop-shaping, root locus to $H_2$ or $H_\infty$ norm based optimal controllers. They provide controllers which are mathematically more rigorous, we are more certain in terms of the boundedness of the output for a known ($H_2$) or even unknown ($H_\infty$) disturbance and noise signals. 

Robust control basically achieves these by several facets, key of which lies in constructing the linear matrix inequalities \cite{DnP13} or the algebraic riccati equations \cite{ZDG96} which put forward the same ideas of stability (e.g. lyapunov) that we have known. The basis of the $H_\infty$ optimal controller is how low can the sensitivity transfer function be pulled across all frequencies. This guarantees that no matter what the noise is or what its frequency content is, we are always assured some performance and stability guarantees. 

The weighting functions are an important part of the robust control framework. It is a design technique to translate the knowledge of classical loop shaping methods into the $H_\infty$ norm minimization based controller synthesis. There are several ways to think of the weighting functions. One could think of the weighting functions as the bounds on the sensitivity and complementary sensitivity functions ch 7 \cite{skogestad}. a very unique way to design the weighting functions, other than the bode based approach is the nyquist approach where the weights are construed as frequency dependent loci of distance from the -1 point in nyquist plane \cite{BMS96,tannenbaum}.

The paper is structured as follows: In Section~\ref{sec:problem statement}, we define the plant which we want to regulate and we try to motivate the readers why is it significant to the power electronics applications. We also introduce the readers to the weighting function based synthesis of robust controllers. We show the mathematics of norm minimization, and give a generalized robust framework encompassing most control applications. Once the plant and controller is formulated, we apply those concepts in Section ~\ref{sec:robustdesign}. Here we cover the dc/ac applications for two classes of grid forming inverters based on their orders of plant. Section~\ref{sec:Bandwidth} establishes a one-to-one correspondence between the classical inner loop-outer loop architecture and controllers synthesized from robust control frameworks. It is shown that when the robust synthesis tool \textbf{hinfsyn} gets the exact same inputs as the classical controller (all the feedforward and feedback signals), robust control synthesizes the exact same \textit{structure} as the multiloop structure. This part of the work is not novel and has been already published in~\cite{Johnson2015OptimalSF}. Our approach differs from~\cite{Johnson2015OptimalSF} in extending this claim to a more generic scenario. We do not derive explicit transfer functions but formulate the plant in state space form, which makes changes in plant structure (addition of filter components or disturbances) easier. Finally we come to the main contribution of the paper where we try to discover the hidden layers within the machinations of robust control and try to find a better bound than the conventional thumb-rule where we assume the inner loop has to be ten times faster than the outer loop.  

\section{Plant and Weighting function based $H_\infty$ controller synthesis}
\label{sec:problem statement}
In this section we will give a scope of the project and delineate the plant and the control formulation we are aiming for. We will begin with a generalized discussion of the scope of robust control formulation for minimizing the $H_\infty$ norm. We will look into the multifarious use of the robust control, and how the classical control formulation can be converted to its equivalents in the robust control framework.

\subsection{The Plant}
\label{sec:PI}
There are several models where we have verified the robust control formulations to. These models differ by the following (1) the filters applied (2) the implementation, switched or averaged model.

\begin{figure}[t]
\vspace{-0pt}
  \begin{center}
  
     \includegraphics[width=0.5\textwidth]{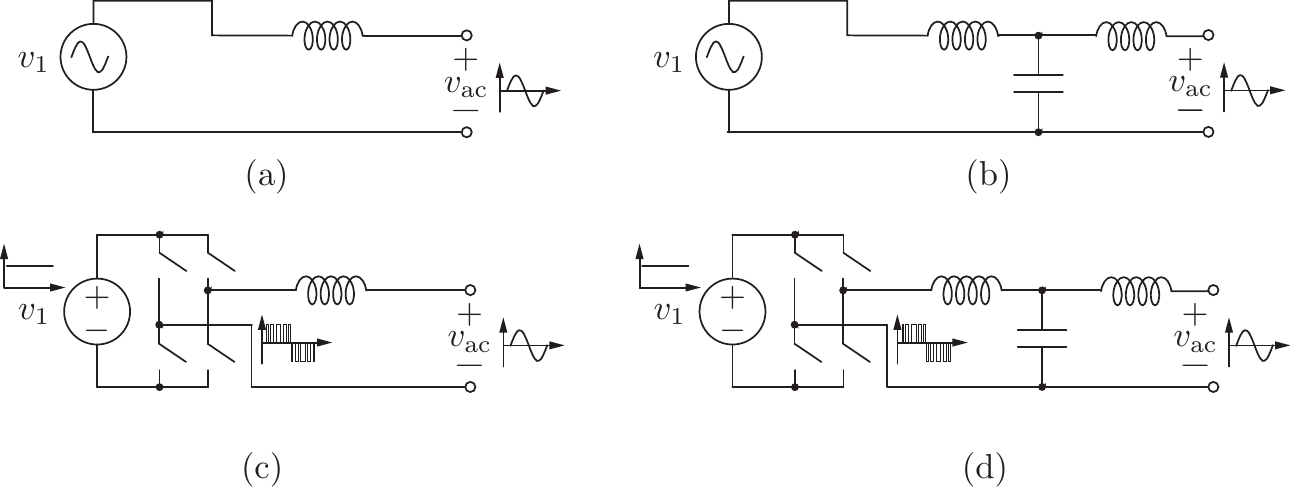}
  \end{center}
  \vspace{-10pt}
  \caption{Figure showing the different inverter configurations that will serve as models}  
  \vspace{-15pt}
  \label{fig:inverterconfig}
\end{figure}
From the Fig. \ref{fig:inverterconfig}, we now focus on the different plant configurations. Fig.  \ref{fig:inverterconfig} (a) is one of the simplest and most intuitive plant. In such a case, the voltage $v_\text{ac}$ is known to us and is an exogenous input, a disturbance.  The voltage on the left, $v_1$ is the controllable voltage. The controlled quantity is the inductor current. We will measure the current flowing in between the two voltages and compare it to a reference quantity and pass the input through a controller. The controller output is directly fed to the voltage source $v_1$. Thus we observe that $v_1$ in Fig. \ref{fig:inverterconfig} (a),(b) is a dependent voltage source. As the reference is changed or the disturbance, $v_\text{ac}$ is changed, the controller updates the $v_1$ to ensure that the reference current is tracked.

Now consider Fig. \ref{fig:inverterconfig}(c). This is more close to the true physical hardware. In this, the ubiquitous concept of pulse-width modulation (PWM) is used to generate the ac waveform from a dc source, ($v_1$). The inverter has two input and two output terminals. Across the input terminals, we connect a dc source $v_1$, which could be a battery or PV. Across the output terminal we have the filter connected in series with the voltage, $v_\text{ac}$, which could be the grid or a simple resistor. A generic treatment of PWM converters can be found in \cite{lipo03}. The PWM output is shown at the output of inverter in Fig. \ref{fig:inverterconfig}(c), (d). There are observable sharp pulses which result in a very highly noisy ac waveform. The 50/60 Hz component is hidden as an average beneath this switched waveform and would be visible only when we aggressively filter out the high frequency switching components. The inductor precisely does this. The poorer the filter is, the lower will be the signal-to-noise ratio. Power engineers characterize SNR by a more suitable term called, THD. Intuitively, it is the ratio of the higher switching frequency components(and its multiples) in the grid current to the fundamental component. Grid operators want this THD to be as low as possible to ensure less polluted voltage and current waveforms. This in turn ensure better operation of the appliances connected to this grid since they will be subjected to less noise and heating (higher order harmonics cause useless heating which is harmful for the appliances).
 \begin{figure}[t]
  \begin{center}
     \includegraphics[width=0.4\textwidth]{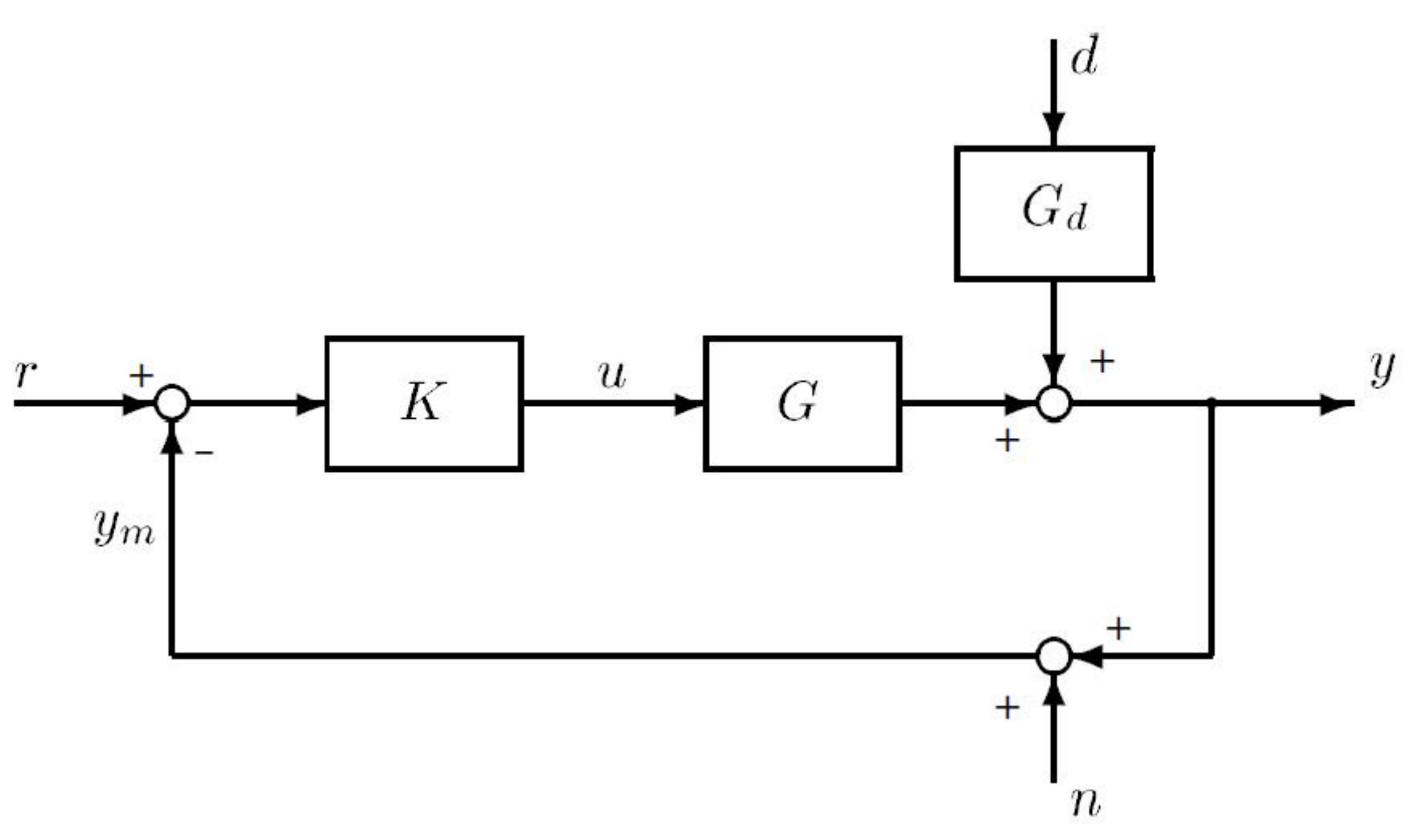}
  \end{center}
  \vspace{-20pt}
  \caption{Figure showing basic control diagram as in ~\cite{skogestad}}  
  \vspace{-15pt}
  \label{fig:basicctrl}
\end{figure}
It has been shown that the single L filter is not alone capable of contributing much to decreasing the THD. Increasing the order of filter to a $LC$ or a $LCL$ is much more beneficial, since they offer better filtering at net lower const of filter investment. \cite{vjohn10} does an optimization to show that LCL filters are most optimized to obtain best THD performance for grid connected inverters.

Having set a good motivation for the different filters and models, we shall now move our focus to what could be the generalized contribution of robust control in our project.

\subsection{Weighting function based $H_\infty$ controller synthesis}
The generalized plant that we come across too often in SISO control is shown as follows. If we define the error as, $e=r-y_m$, then by the robust control formulation, we need to minimized the $H_\infty$ norm of this error to all the exogenous inputs.

We first relate the error to the other exogenous inputs as follows,
 \begin{flalign}
         E(s)= \frac{1}{1+GK}\,R(s)+ \frac{G_d}{1+GK}\,D(s)+ \frac{GK}{1+GK}\,N(s)
    \end{flalign}
    The control nomenclature gives us two interesting parameters, namely the Sensitivity, $S(s)$, and Complementary Sensitivity, $T(s)$, transfer functions. Mathematically we define these two quantities as,
      \begin{flalign*}
       S=\frac{1}{1+GK}\quad T=\frac{GK}{1+GK}
    \end{flalign*}   
    Any standard control literature has tons of description of each of these terms and they form the cornerstone of entire classical control. It turns out that they are equally important for the robust control framework.
     \begin{figure}[t]
  \begin{center}
     \includegraphics[width=0.45\textwidth]{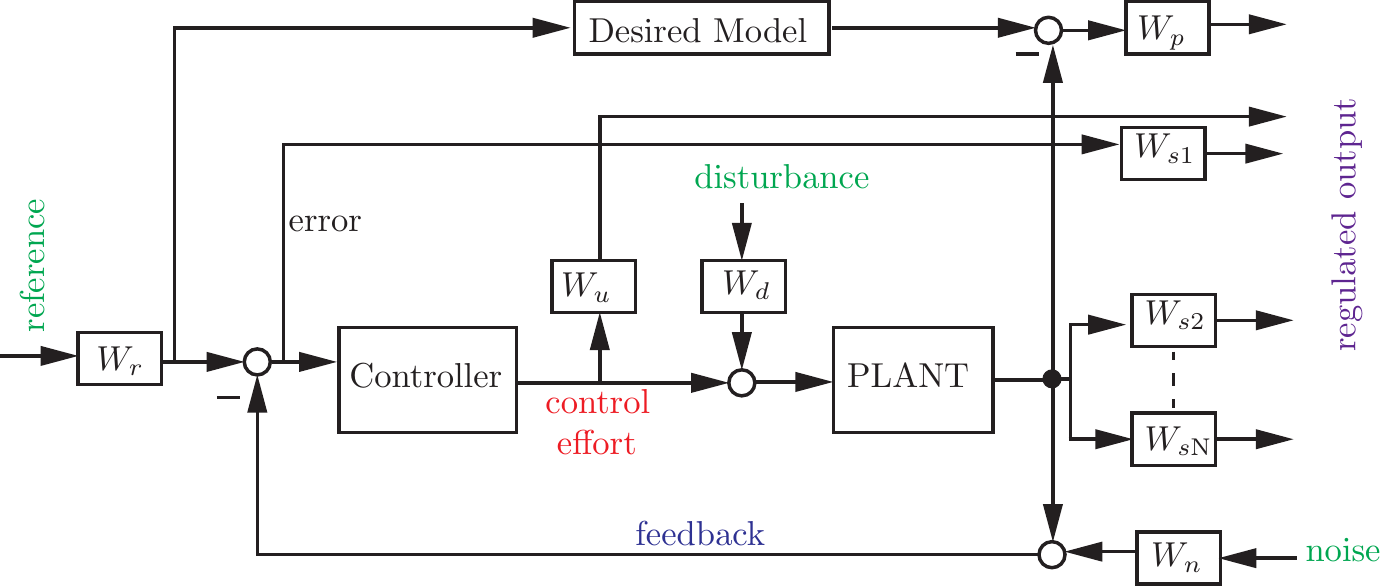}
  \end{center}
  \vspace{-10pt}
  \caption{Figure showing generalized $H_\infty$ control diagram}  
  \vspace{-10pt}
  \label{fig:generalizedhinf}
\end{figure}
    We will need to achieve the following control objective
       \begin{flalign}
      \text{minimize}  \Bigg\Vert  \begin{bmatrix}e(s)/r(s)\\e(s)/d(s)\\e(s)/n(s)\end{bmatrix}\Bigg\Vert^2_{\infty} =   \text{minimize}  \Bigg\Vert  \begin{bmatrix}S(s)\\G_d(s)S(s)\\T(s)\end{bmatrix}\Bigg\Vert^2_{\infty}
      \label{eq:normmin}
    \end{flalign}  

This means we will have to formulate a controller that will minimize the above expression in the $H_\infty$ norm sense. That is it will minimize the maximum value of $\|S^2+(G_dS)^2+T^2\|$ at any frequencies. We can solve to find a controller that satisfies this for the output feedback case by either solving the LMI \cite{DnP13} or solve the algebraic Riccati equation (ARE) \cite{ZDG96} or use MATLAB's internal \textbf{hinfsyn} solvers.

In a previous assignment, a code was developed to solve the same problem using both MATLAB's internal \textbf{hinfsyn} and the ARE of \cite{ZDG96} (formulate it as ARE and then use \textbf{icare} to solve the ARE) yielded exactly same controller. Gaining motivation from that, we continue to formulate the complex systems and problems into a form that the \textbf{hinfsyn} more readily realizes and solves. In the next section we deal with how to formulate the control problem into a form that \textbf{hinfsyn} realizes and solves.

\subsubsection{Study of Generalized $H_\infty$ Control Framework}

Let us now look at what arranging the control problem in the form of Fig \ref{fig:generalizedhinf} helps achieve. The primary objective of the paper is that we want to reduce all the regulated variables' with respect to any exogenous input which is bounded to 1. This means, for any $r,d,n<1$ (where $r,d,n$ refer to reference, disturbance and noise respectively) all the regulated variables will have values which are negligible. 

From the discussion of \eqref{eq:normmin}, we understand that we need to minimized the error w.r.t. $d,r,n$. But what if we slightly changed this objective, and rather want to minimize $eW_\text{s1}$ for any $r,d,n<1$. But then, the question remains, how does it relate to \eqref{eq:normmin}, and even if it does, what is $W_\text{s1}$?

 \begin{figure}[t]
  \begin{center}
     \includegraphics[width=0.5\textwidth]{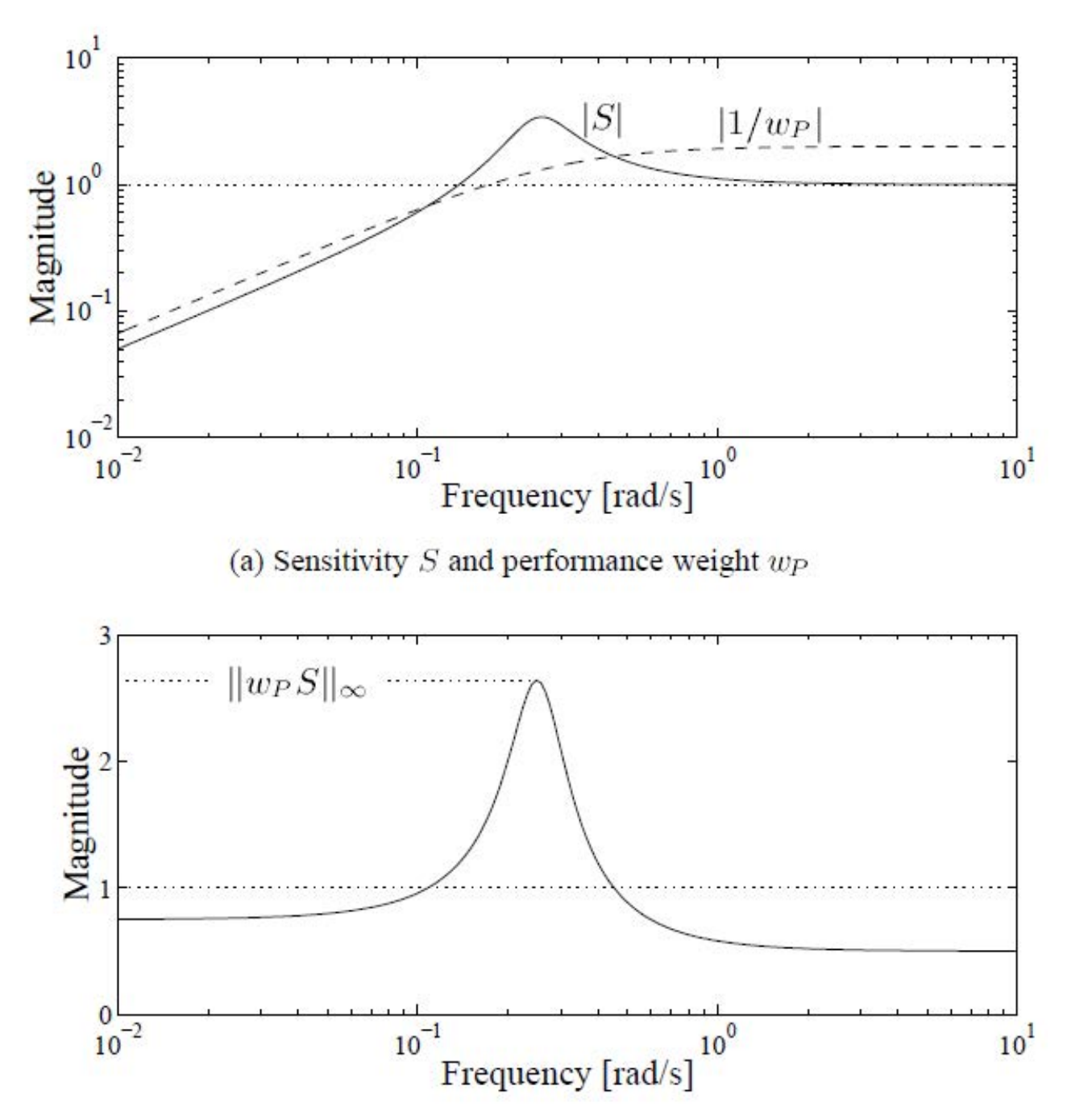}
  \end{center}
  \vspace{-20pt}
  \caption{Figure showing how to select weighing function for error, $W_\text{s1}$ as in~\cite{ZDG96}}  
  \vspace{0pt}
  \label{fig:Ws1selection}
\end{figure}

Fig. \ref{fig:Ws1selection} (a) a hypothetical $S$ that might come up for a standard system. We might want the $S$ to look like that depending upon, the specification of reference tracking, disturbance rejection and noise rejection. We then construct a function $1/w_p$ such that at most of the frequencies, it is larger than $S$. Thus the product of $w_pS$ would be less than 1 at most of the frequencies. The robust control formulation would basically aim to reduce the peak of $\|w_pS\|_\infty$. In our formulation of Fig. \ref{fig:generalizedhinf}, we refer to the same function $w_p$ and $W_\text{s1}$. Thus, when we solve the \textbf{hinfysyn}, the output $\gamma$ is actually the value, $\|w_pS\|_\infty$ or , $\|W_\text{s1}S\|_\infty$.

A similar discussion can be followed when we want to minimize control effort usage. This is similar to LQR when we want to penalize high control effort. If we relate the error to reference as $e(s)=Sr(s)$, similarly the control effort, $u$, is related as $u=Ke=KSr(s)$. We design a weighing function and the argument is exactly the same as the design of Fig. \ref{fig:Ws1selection}. We keep the gain of $W_u$ high at low frequencies and low at higher frequencies. This is because we want $u$ to be of very small values at high frequencies since the control effort would mainly be at low frequencies. Thus we can see $W_u$ as penalizing factor.
\subsubsection{Design of the weighting functions}
We could understand the full potential of weighting functions, we can appreciate the following uses other than the reduction of control effort and errors.

In some applications like flight modelling, experts often have excellent reference models which they want the plant and controller to emulate. This is a rare scenario when one has an exact idea of how the desired model should look at all frequency. In such a case, as we see in Fig \ref{fig:generalizedhinf}, the output of the actual plant is compared to an output of the desired model which we want the plant+controller to emulate. Both, get the same inputs and we want to minimize the error between the outputs. This can also be a potential specification for the controller.
 \begin{figure}[b]
 \vspace{-10pt}
\begin{minipage}[b]{0.45\linewidth}
\centering
\includegraphics[width=\textwidth]{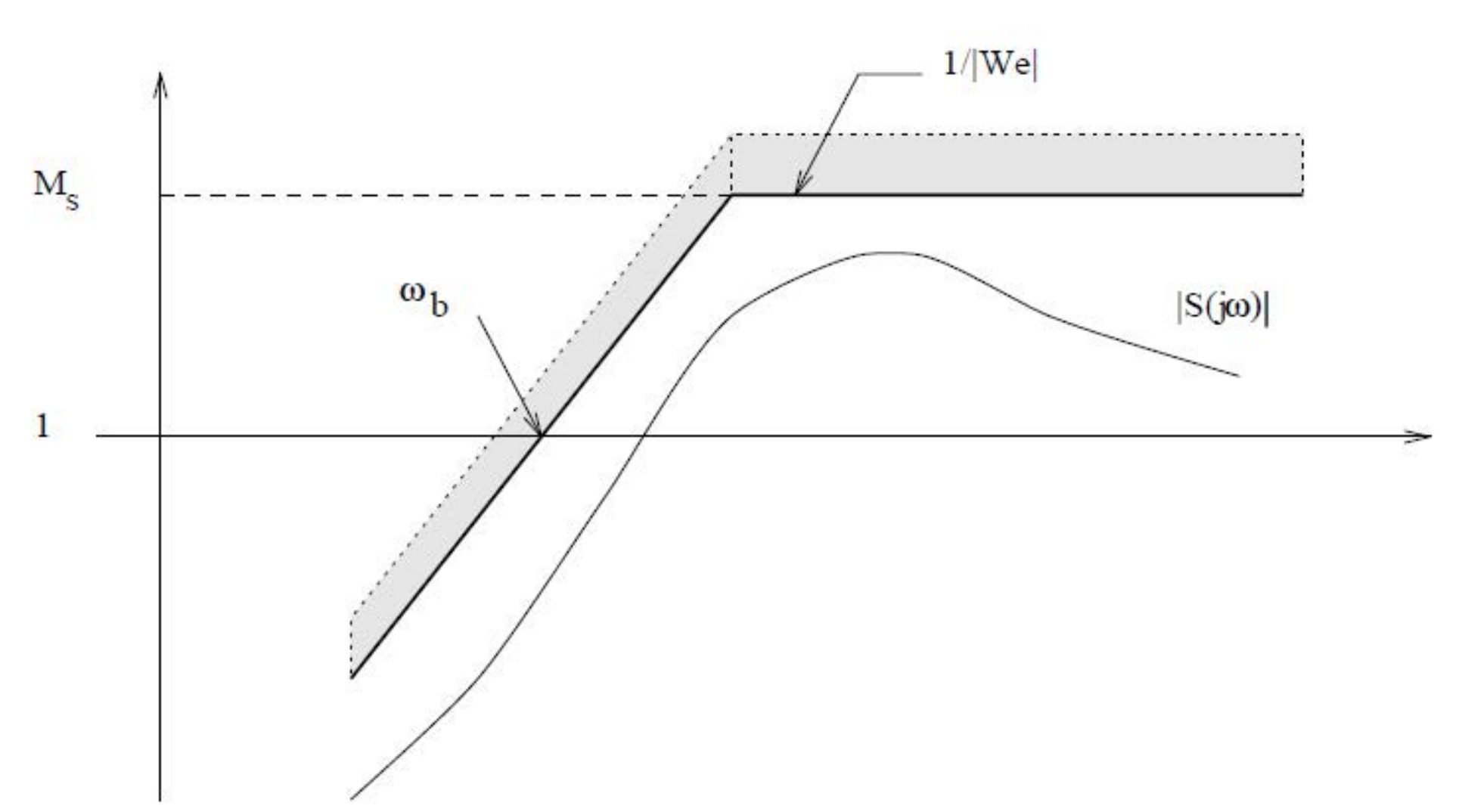}
 \caption{Structure of generalized error weighing function, $W_\text{s1}$~\cite{skogestad}}  
\end{minipage}
\begin{minipage}[b]{0.45\linewidth}
\centering
\includegraphics[width=\textwidth]{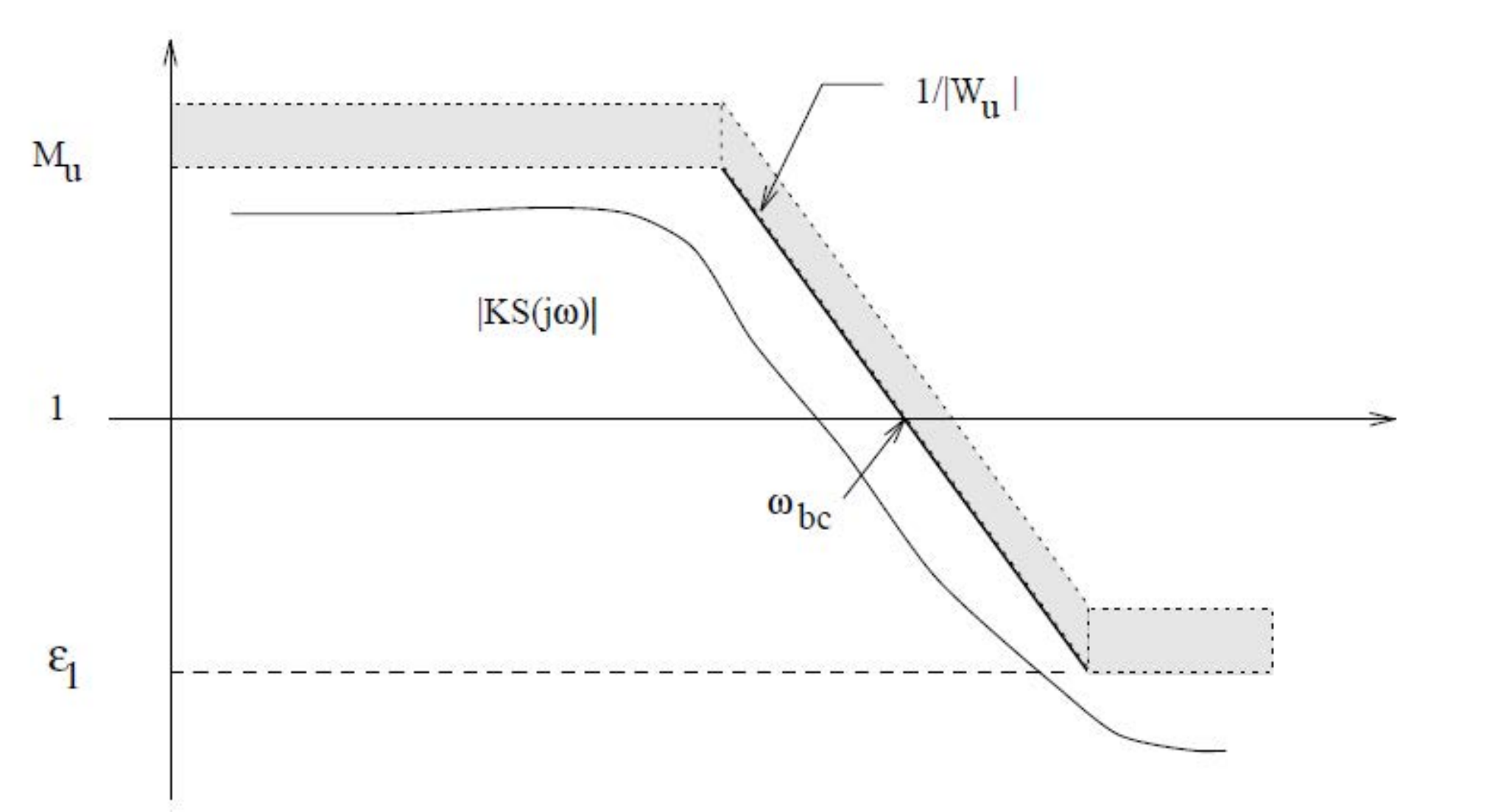}
 \caption{Structure of generalized error weighing function, $W_\text{u}$~\cite{skogestad}}  
\end{minipage}
 \end{figure}
The inputs are supposed to be normalized and their weighting functions like $W_d,W_r,W_n$ have a slightly different perspective. These are generally normalization factors. So if we expect the maximum value of the disturbance to be $M$, then $W_d=1/M$. The argument is, if the controller can minimize the norm of the regulated variable to a disturbance of value $M$, to $\gamma$, then for any value of disturbance less than $M$, the maximum norm of regulated variable will be less than $\gamma$.

We could, similarly conjure up an infinite number of such regulated-variables and then force the controller to satisfy a norm minimization. 

\subsubsection{Caveat of the weighting functions}

The most obvious caveat is to construct some weighting function blindly, not knowing what it exactly means or to use pure trial and error. One must understand that for one-degree of freedom controller, where the \textbf{hinfsyn} will only generate the controller K, we only have limited flexibility that only one specification can be given at one particular frequency. In theory, we can construct multiple regulated outputs and design weighting functions for each. But how these different weighting functions conflict and upset the design is important. An easy thing to look out for when iterating different weighting function is to check the value of $\gamma$. If that is too high, it means the weighting functions are imposing absurd demands on the controller structure.

 \section{Controller Synthesis and Simulation Results}\label{sec:robustdesign}
In this section we will focus on two sample applications of the control theory that we developed in the previous section. We will discuss the weighting function designs and conclude with intuitions gathered from the results based on simulations with the controllers we synthesized.
\subsection{DC current control of a L plant}
\begin{figure}[t]
\vspace{-10pt}
  \begin{center}
     \includegraphics[width=0.4\textwidth]{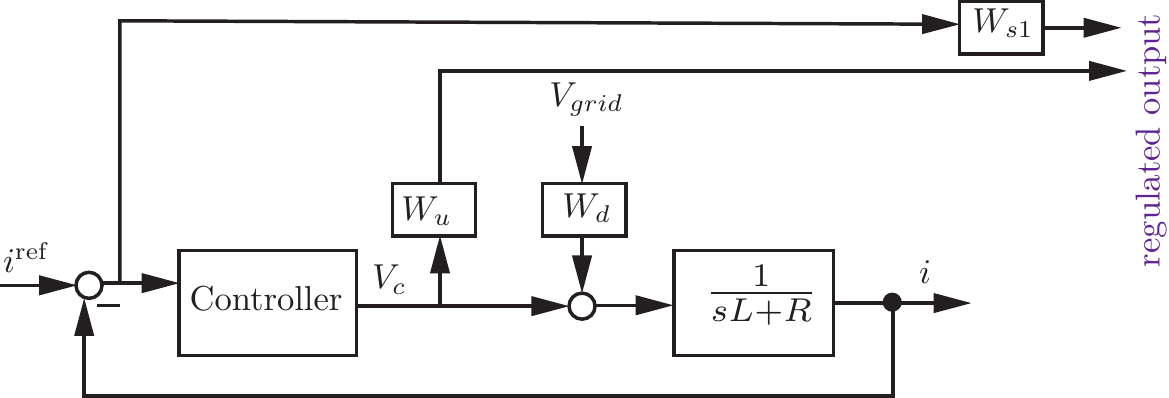}
  \end{center}
    \vspace{-10pt}
  \caption{L Filter based inverter current control}  
  \label{fig:LinverterBlock}
\end{figure}
The Fig. \ref{fig:generalizedhinf} can be simplified to Fig. \ref{fig:LinverterBlock}. To design the weighting functions, $W_d$, $W_u$, $W_\text{s1}$, we design them as the following.
\begin{flalign}
\label{eq:Ws1TF}
W_\text{s1}&=\frac{s/M_s+\omega_b}{s+\epsilon \omega_b}\\
W_u&=\frac{s+\omega_{bc}/M_u}{\epsilon_1 s+\omega_{bc}}\\\label{eq:WdTF}
W_d&=\frac{1}{V_\text{ac}^\text{max}}
\end{flalign}

The value of $\omega_b$ puts an lower bound on the open loop bandwidth, while $\omega_\text{bc}$ is an upper bound on the open loop bandwidth. $M$ is selected to be around 2. $\epsilon,\epsilon_1$ are both selected to be very small numbers in the range of $10^{-3}$ or smaller. The reason they have to exist is because \textbf{hinfsyn} cannot handle singularities like an integrator or differentiator. It has to be cut off to dc asymptote at low frequencies. This also makes sense because we often do that to avoid saturation and wind-up. The following figures would be better visual representation of the claims.

 To improve upon the controller performance, one could go for higher order transfer functions for each of the weighing functions. The controller thus synthesized would also grow up in orders, thus giving better performance. We would however end up paying more cost in terms of control effort, $u$.
 
\begin{figure}[b]
 \vspace{-10pt}
  \begin{center}
     \includegraphics[width=0.5\textwidth]{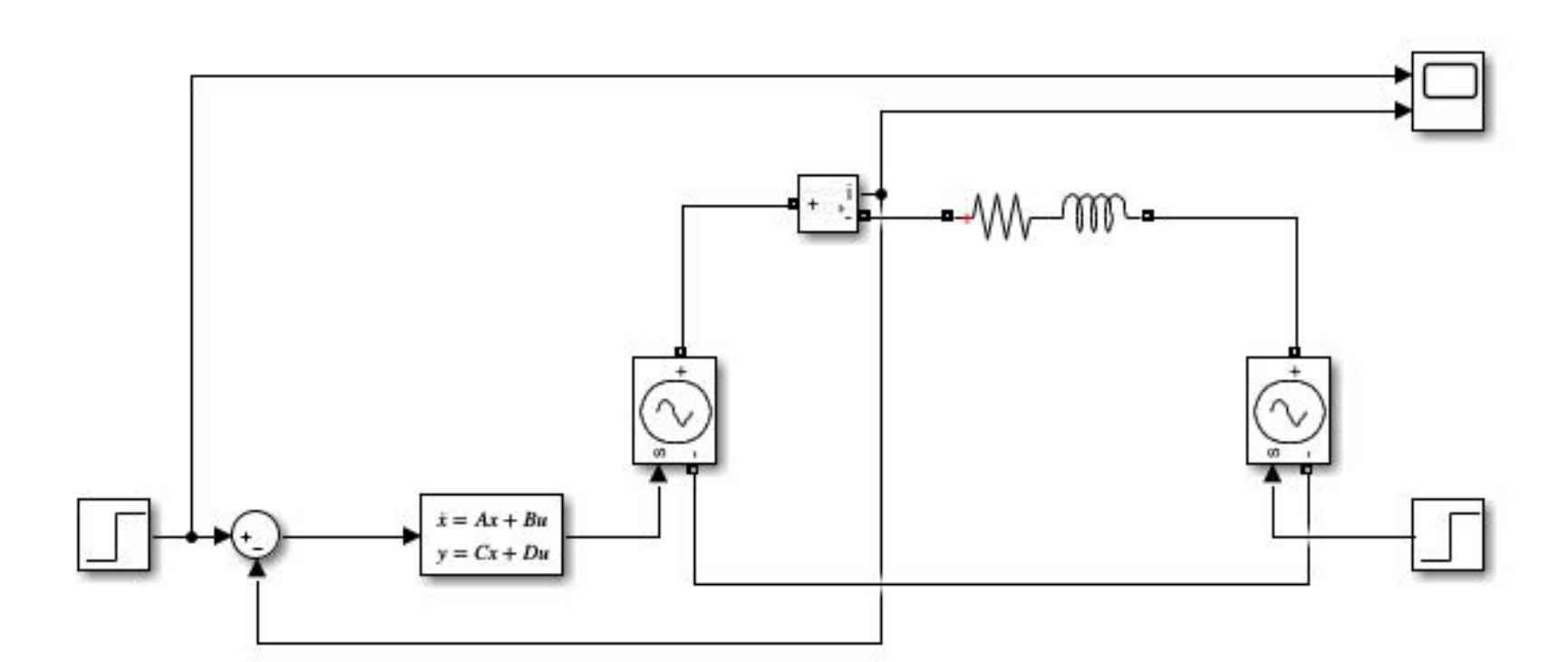}
  \end{center}
    \vspace{-20pt}
  \caption{MATLAB simulink model for L filter}  
  \label{fig:simLfilter}
\end{figure}

 Let us now consider the simulation in SIMULINK, as shown in Fig. \ref{fig:simLfilter}. The premise has been set up in the discussion of Fig. \ref{fig:inverterconfig} (a). The only small change we do here is, we make everything dc for simplicity. We assume the disturbance, inductor current and the control voltage all to be dc quantities.

We came up with three designs for the controller shown in Fig. \ref{fig:simLfilter}. The controllers are as follows (1) PI controller (2) $H_\infty$ controller based on single order $W_\text{s1}$ (3)  $H_\infty$ controller based on second order $W_\text{s1}$.

 \begin{figure}[t]
 \vspace{-10pt}
  \begin{center}
     \includegraphics[width=0.5\textwidth]{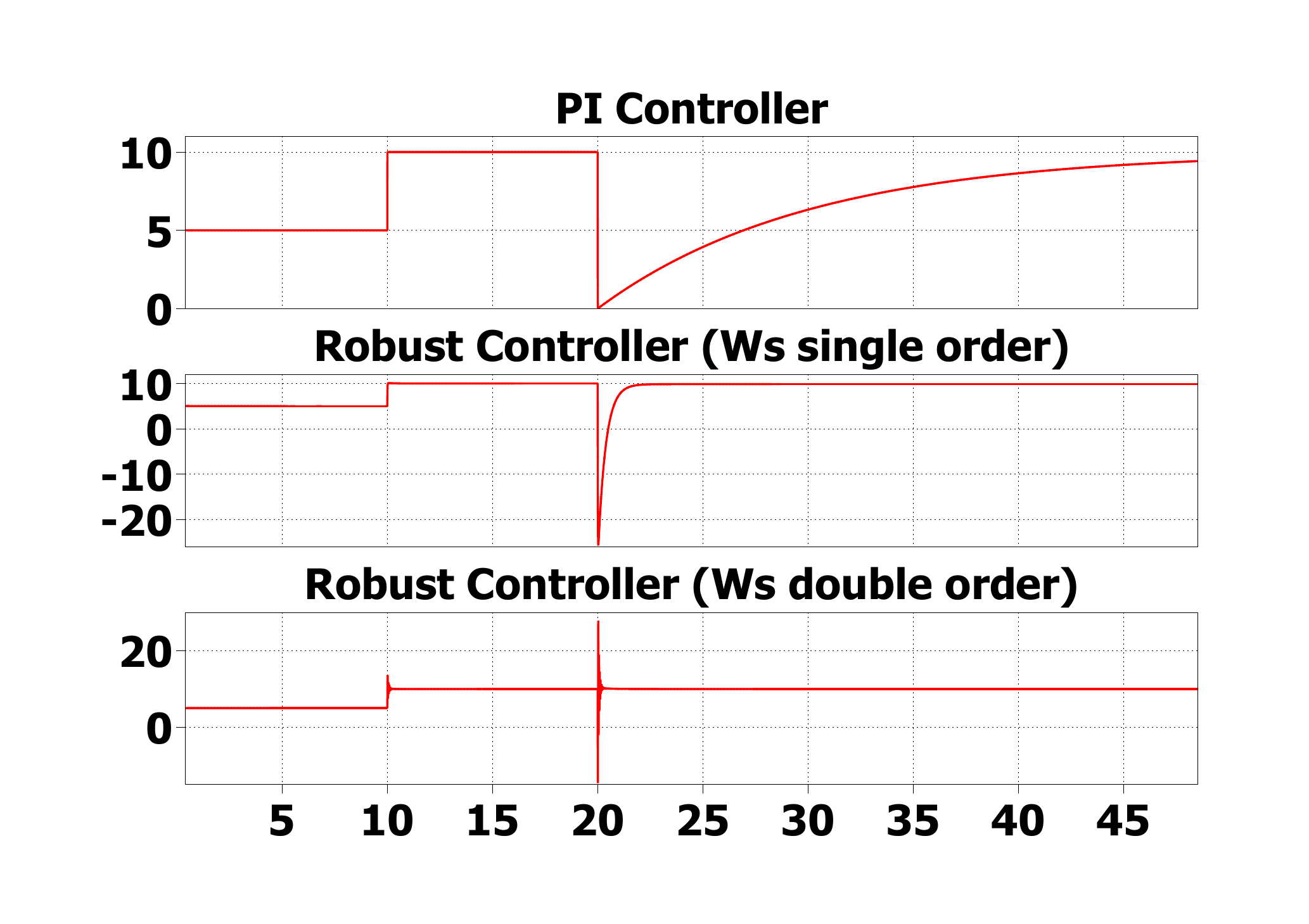}
  \end{center}
  \vspace{-20pt}
  \caption{Figure showing the response of the different types of controllers for the dc current tracking in a L filter}  
  \vspace{-20pt}
  \label{fig:compearePIWs}
\end{figure}

The comparison between the three controllers had to be based on something that was kept constant. Bandwidth seemed the obvious choice since we cannot compare the PI controller with poorer bandwidth to a robust controller having much higher bandwidth. The latter would of course perform better.

The PI controller has been design by inverse-based design principle \cite{skogestad}. If we define $G_c$ as the controller and $G_p$ as the plant transfer function, we can design the controller as, $G_c=(\omega_\text{bw}/s)G_p^{-1}$. In the preceding equation, $\omega_{bw}$ is the desired bandwidth. The resulting controller is a PI controller for  a first order plant transfer function. For fair comparison, we design the weighting function $W_\text{s1}$ by the \eqref{eq:Ws1TF} as follows, $\omega_b=100$Hz, $M_s=2$,$\epsilon=10^{-3}$. We do not have any specification for the $W_u$. For the third case, we cascade the $W_\text{s1}$ to itself to make it a second order. 

The experiment has been set up such that at $t=0 s$, the current reference is 5A, followed by the reference changing to 10 A at $t=10 s$. Finally with this same current reference, we subject the plant to a disturbance by changing the grgid voltage from 0V to 120 V at $t=20 s$.

Fig. \ref{fig:compearePIWs} shows the performance of each of the plant. The PI controller has zero steady state error but the rise time is much longer for the case where disturbance comes in. The robust based controllers have the disturbance rejection embedded into them by employing \eqref{eq:WdTF}. That is the reason their performance to disturbance rejection is much better. We also observe, if we zoom very closely around the steady state values that there is a very small error,  in the order of mA in tracking. This is also expected becasue the robust controllers cannot synthesize perfect integrators. They have a low frequency pole embedded in the PI controller to avoid saturation. Thus, the gain at DC is not infinite but a certain number, reciprocal of $\epsilon$. And this shows up as the steady state error. The error however is negligible for practical purposes.

Also important is to see that the double order $W_{s1}$ has much higher osccilations and overshoots because of the higher dynamics of control effort. It is probably unfair to compare PI controller with the second or $W_{s1}$ based $H_\infty$ controller since the latter generates a second order controller. Having specifications on $W_u$ or $W_d$ coudl increase the order of controller by each order of these respective transfer functions. And this is, I believe the single most important contribution of $H_\infty$ based controller in robust control. No classical techniques would give us a theoretical framework for a controller of order higher than order 1 for a single order plant. The higher order of controller signifies more flexibility at choosing gain at each frequency rathe than simply keeping it a constant or rolling it off at -20 dB, and similar approaches which form the cornerstone of classical control design.
But then one must also acknowledge that since in this, we had $W_u=1$, or basically no constraint on the control effort, our second order $W_{s1}$ based $H_\infty$ controller performs much poorly in terms of the control effort usage, both in terms of peak $u$ as well as oscillations in it.
\vspace{-5pt}
\subsection{AC current control of a LCL plant in Grid Following Inverter}
The motivation of the control comes from the Fig. \ref{fig:inverterconfig} (b). In this, the inductor current, the disturbance voltage, $v_\text{ac}$ and the controlled voltage source are all ac quantities of frequency, $\omega_r$ rad/s which is usually 50 or 60 Hz. It is grid following since there is grid on the other side which maintains the stiff frequency, $\omega_r$. So we can only follow that grid. 
\begin{figure}[t]
 \vspace{-10pt}
  \begin{center}
     \includegraphics[width=0.5\textwidth]{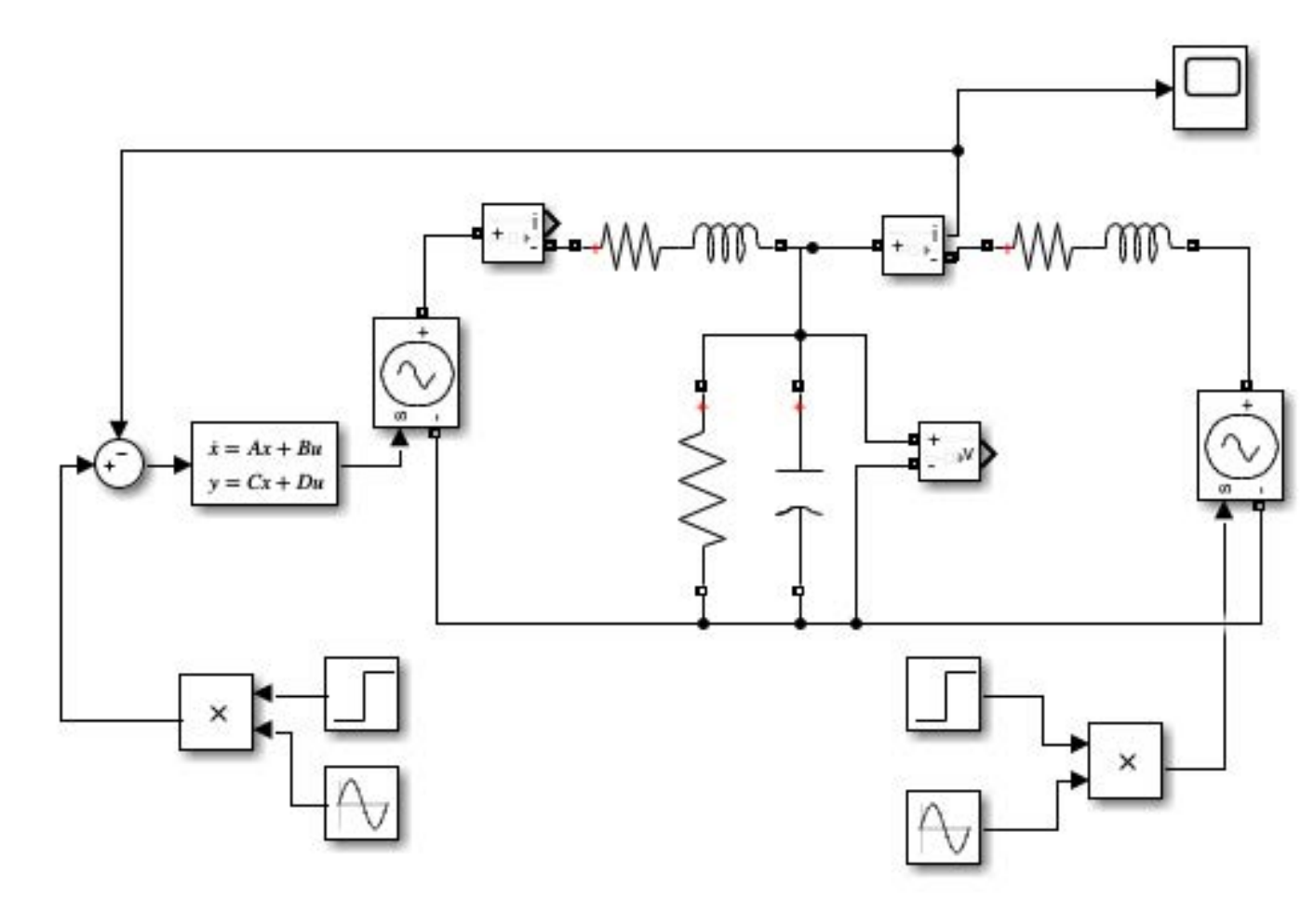}
  \end{center}
  \vspace{-10pt}
  \caption{Figure showing the simulation of a LCL model with AC voltages and current}  
  \vspace{-10pt}
  \label{fig:simLCL}
\end{figure}\\

The design of the weighing functions is slightly more involved in this case. The dc tracking or disturbance rejection in time domain translates to the requirement of infinite open loop gain at dc. Similarly, a reference tracking and disturbance rejection at an ac frequency of $\omega_r$ needs the open loop gain to be significantly large at that frequency.

\begin{figure}[h]
  \begin{center}
     \includegraphics[width=0.5\textwidth]{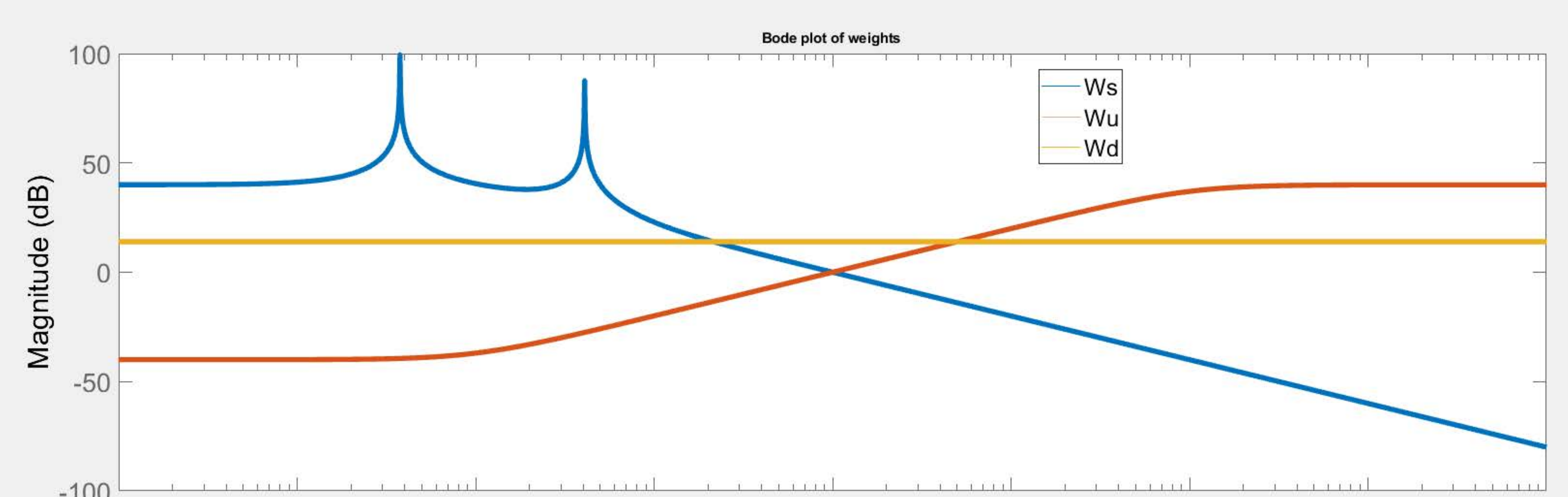}
  \end{center}
  \vspace{-20pt}
  \caption{Figure showing the different weighting functions}  
  \vspace{0pt}
  \label{fig:bodeWsud}
\end{figure}
The following are the weighting functions selected for this 
\begin{flalign*}
W_\text{s1}&=\frac{100}{s/1e3+1}\frac{s^2+2\omega_r+\omega_r^2}{s^2+0.001\,2\omega_r+\omega_r^2}\frac{s^2+2\omega_{res}+\omega_{res}^2}{s^2+0.001\,2\omega_{res}+\omega_{res}^2}\\\vspace{2pt}
W_d&=\frac{1}{v_\text{ac}^\text{max}},\quad\quad W_u=\frac{0.01(s/1e3+1)}{s/1e7+1}
\end{flalign*}
\begin{figure}[b]
  \begin{center}
     \includegraphics[width=0.5\textwidth]{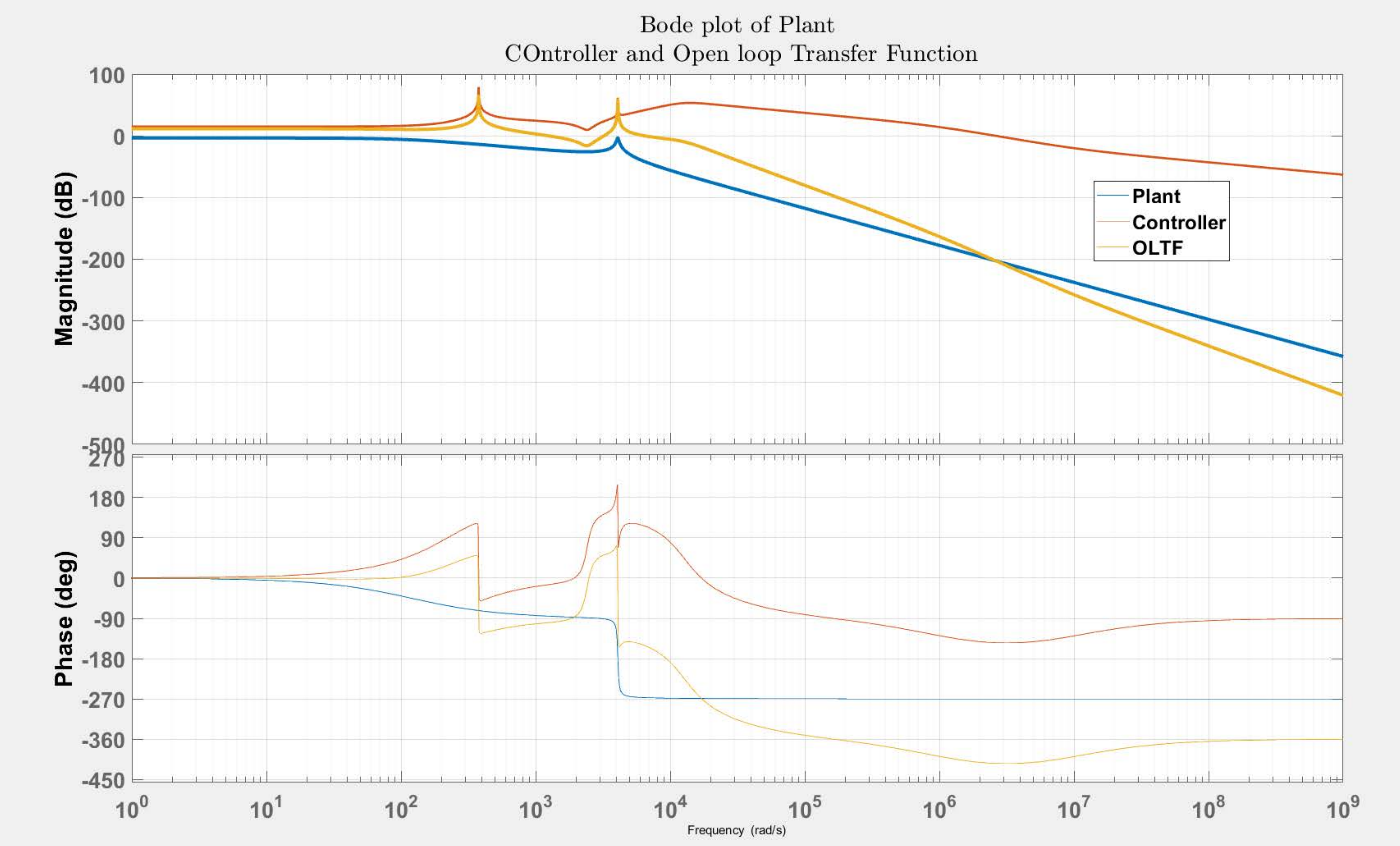}
  \end{center}
  \vspace{-10pt}
  \caption{Figure showing the different bode plots of controller, open loop TF, closed loop TF}  
  \vspace{-15pt}
  \label{fig:bodeallLCL}
\end{figure}
The $W_\text{s1}$ which is responsible for minimizing the error to a reference is shown in blue. It has two peaks. The first peak is at 60 Hz, which is the frequency of both the reference and the disturbance, $\omega_r$. The second peak is slightly involved. The roll-off of $W_\text{s1}$ at higher frequency is needed to make sure that the noise is not amplified. Similarly, the weighting function associated with control effort, $u$, is shown in red. It is penalized at high frequencies by maintaining high gain at high frequencies and low gain at low frequencies. This is mainly done to remove the switching frequency components in the control effort, $u$. This in turn leads to better TDH performance.

The design of $W_d$ is straight forward and borrows from the definition of the previous design. We just use normalization with the maximum peak of the $v_\text{ac}$. 

We notice that the plant has a $LCL$ filter. The parallel combination of the inductor and the capacitor starts resonating at the frequency of $\omega_\text{res}=1/\sqrt{\frac{L_lL_2C}{L_1+L_2}}$. During the step change of reference, the output also undergoes sharp changes. These sharp changes for averaged models or the PWM pulses in the switched model can be decomposed to give infinite frequencies in the fourier spectrum. It so happens that one of these components might have a value of $\omega_\text{res}$ which excited resonance and soon makes the plant unstable. To reduce its effect, we must have another dominant peak at the frequency of the $\omega_\text{res}$ embedded into the design of $W_{s1}$. This is widely popular as the concept of active damping. We could as well have used passive damping, wherein we increase the resistance of the $LCL$ filter but that method to counteract resonance based instability involves a lot of power dissipation in the resistors which is not favourable.

The bode plots in Fig. \ref{fig:bodeallLCL} show the controller (in red), the open loop transfer function (in yellow), and the plant without any controllers (in blue). The plant without any controllers shows the presence of the oscillatory modes at $\omega_\text{res}$. The controller has to simultaneously satisfy the contrasting demands of the two loop shaping functions $W_{s1}$ and $W_u$ at frequencies wheres they intersect in magnitudes above 0 dB (see Fig. \ref{fig:bodeWsud}). This leads to a controller having un-pronounced peak at $\omega_\text{res}$. However, we observe the open loop TF to have a peak at the $\omega_\text{res}$. 

 \begin{figure}[t]
  \begin{center}
     \includegraphics[width=0.5\textwidth]{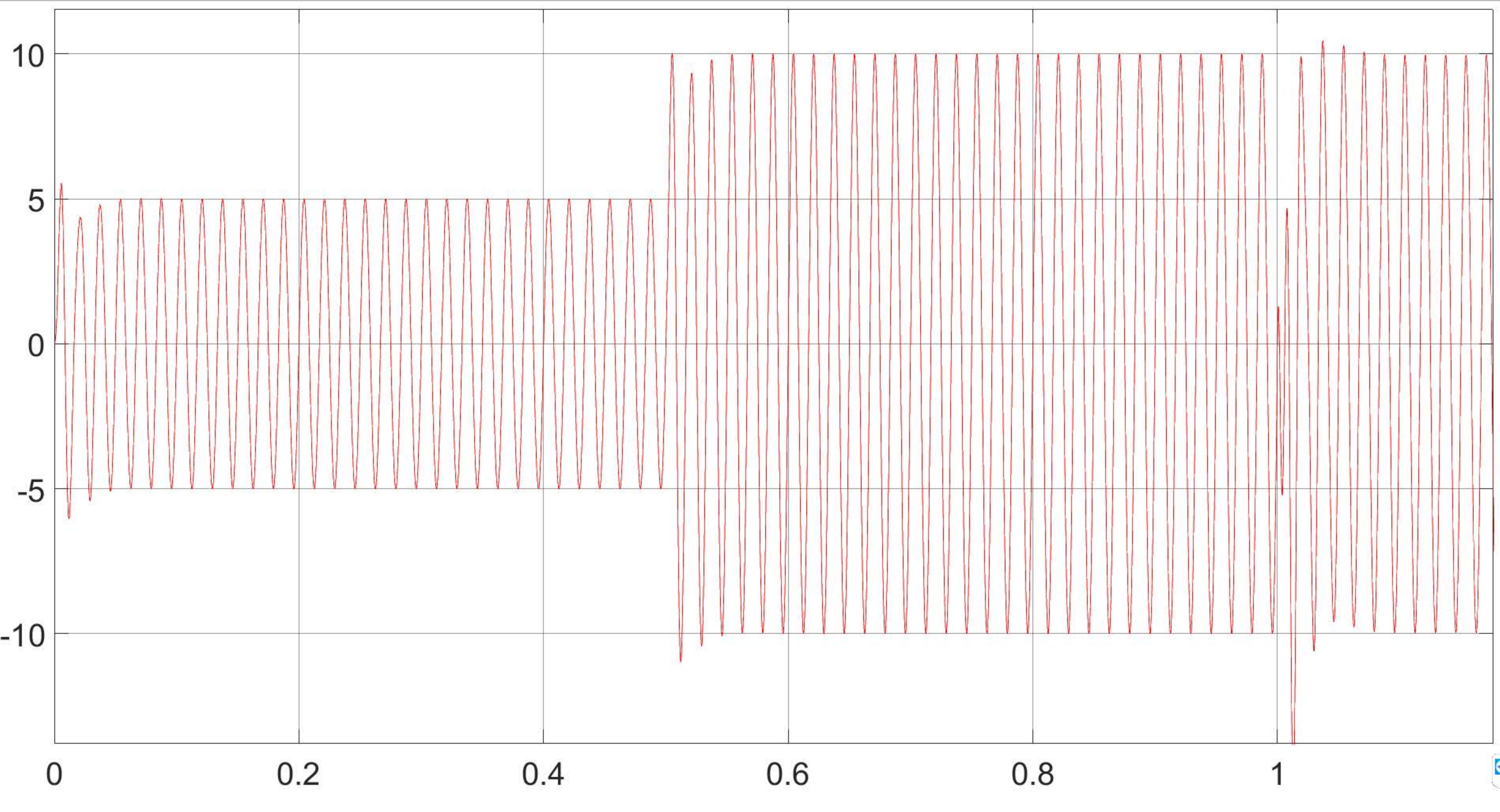}
  \end{center}
  \vspace{-15pt}
  \caption{Figure showing the inductor current response in simulation for the $LCL$ model with ac waveforms}  
  \vspace{-15pt}
  \label{fig:simLCLresult}
\end{figure}
\vspace{-00pt}
\begin{figure}[b]
\vspace{-10pt}
  \begin{center}
     \includegraphics[width=0.5\textwidth]{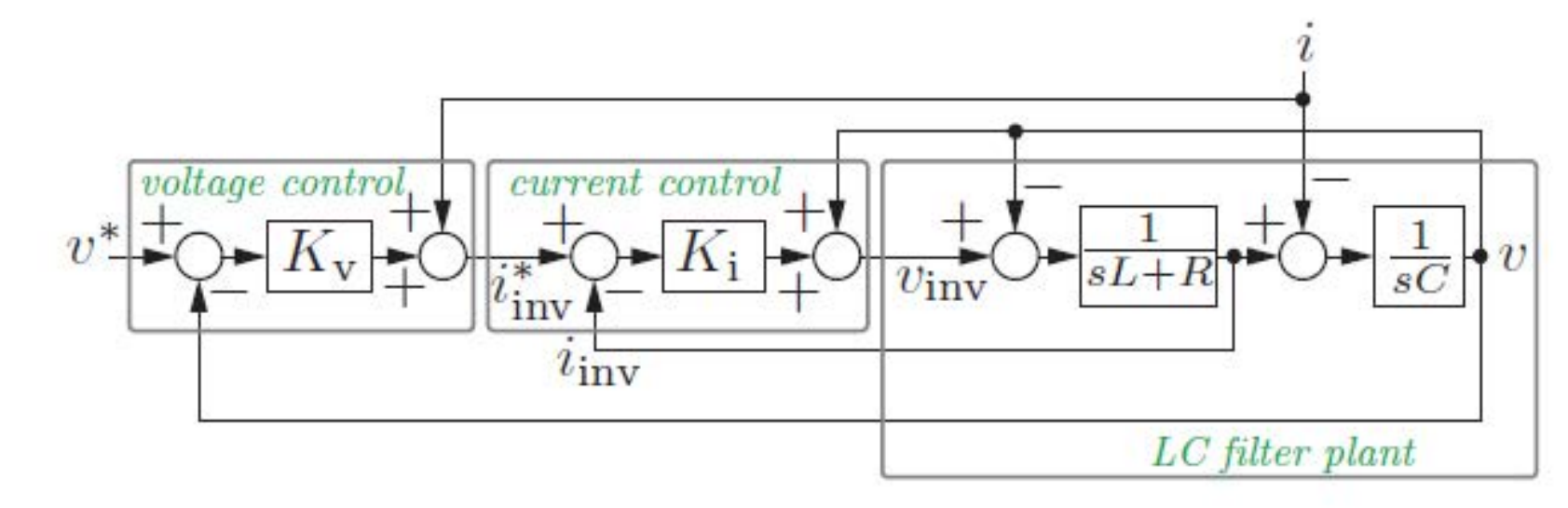}
  \end{center}
  \vspace{-20pt}
  \caption{Figure in~\cite{Johnson2015OptimalSF} showing the inner-current and outer-voltage architecture in inverters}
  \label{fig:ICOVBLKdiag}
  \vspace{0pt}
\end{figure}
 The simulation results in Fig. \ref{fig:simLCLresult} shows that the sinusoidal current is perfectly tracked. We have the following conditions (1) At $t=0 s$, $i^\text{ref}=5A_\text{pk}$ and $v_\text{ac}=0 V_\text{pk}$, (2) At $t=0.5 s$, $i^\text{ref}=10A_\text{pk}$ and $v_\text{ac}=0 V_\text{pk}$, and (3) At $t=0 s$, $i^\text{ref}=10A_\text{pk}$ and $v_\text{ac}=120 V_\text{pk}$. We observe excellent tracking and disturbance rejection within one sub-cycle. 
 \section{Translating classical Multiloop structure to a robust formulation} \label{sec:Bandwidth}
 We all know of control applications when we set the reference for one physical quantity but then there are hidden layers beneath the primary control which controls the secondary and tertiary quantities inside loops. This inner-outer loop control architecture is very popular in applications where the loops and the quantities it controls are separated by their natural time constants or in cases where it has been shown that the inner control gives rise to better transient performance and hence better overall reference tracking of the desired physical quantity. The only assumption behind this control configuration is that the inner control loop needs to be much faster than the outer control loop.
 
 In control of the rotating machines, the outer loop is usually the speed control of the machine and the inner loop involved the control of the current through the coils of the machines. The current controller is dictated by the electrical time constant which is much smaller than the mechanical time constant which in turn determines the speed tracking. This automatically establishes the inner control loop being faster than the outer control loop even without any explicit controller.
 
\begin{figure}[t]
\vspace{-10pt}
  \begin{center}
     \includegraphics[width=0.36\textwidth]{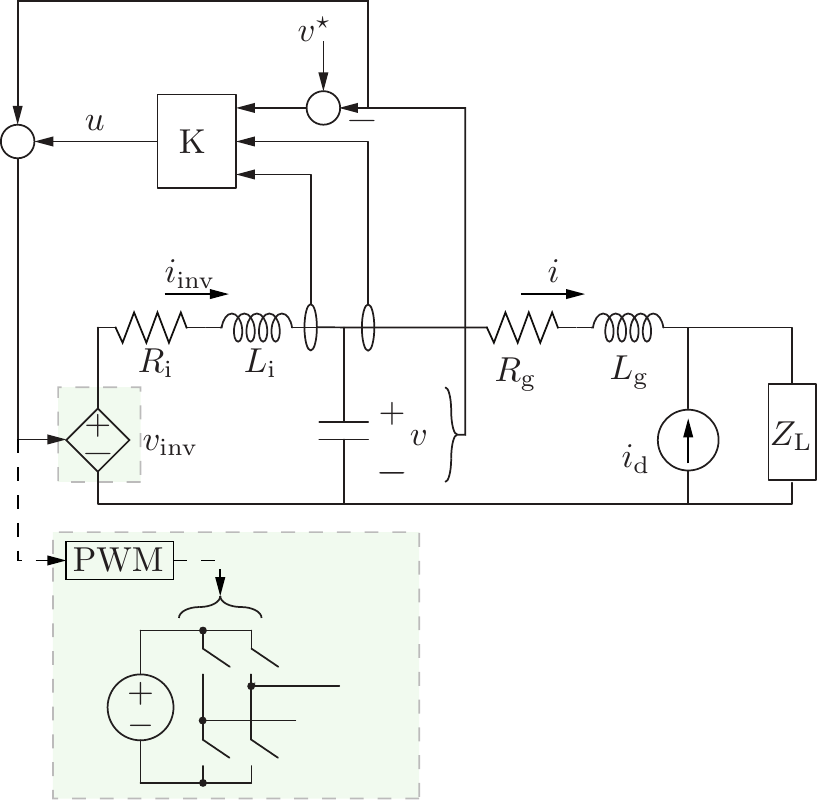}
  \end{center}
  \vspace{-10pt}
  \caption{Figure showing the plant and robust control application} 
  \label{fig:LCLcurrentdist}
  \vspace{-10pt}
  \end{figure}

 In many power converters, the similar thing happens where the outer loop controls the voltage through a capacitor and the inner current loop controls the current through an inductor which then ultimately feeds the capacitor (See Fig. \ref{fig:LCLcurrentdist}, the $i_\text{inv}$ and $v$ for instance). Thus having the inner loop fastens the response of the outer voltage controller. 
  \begin{figure}[b]
  \begin{center}
     \includegraphics[width=0.5 \textwidth]{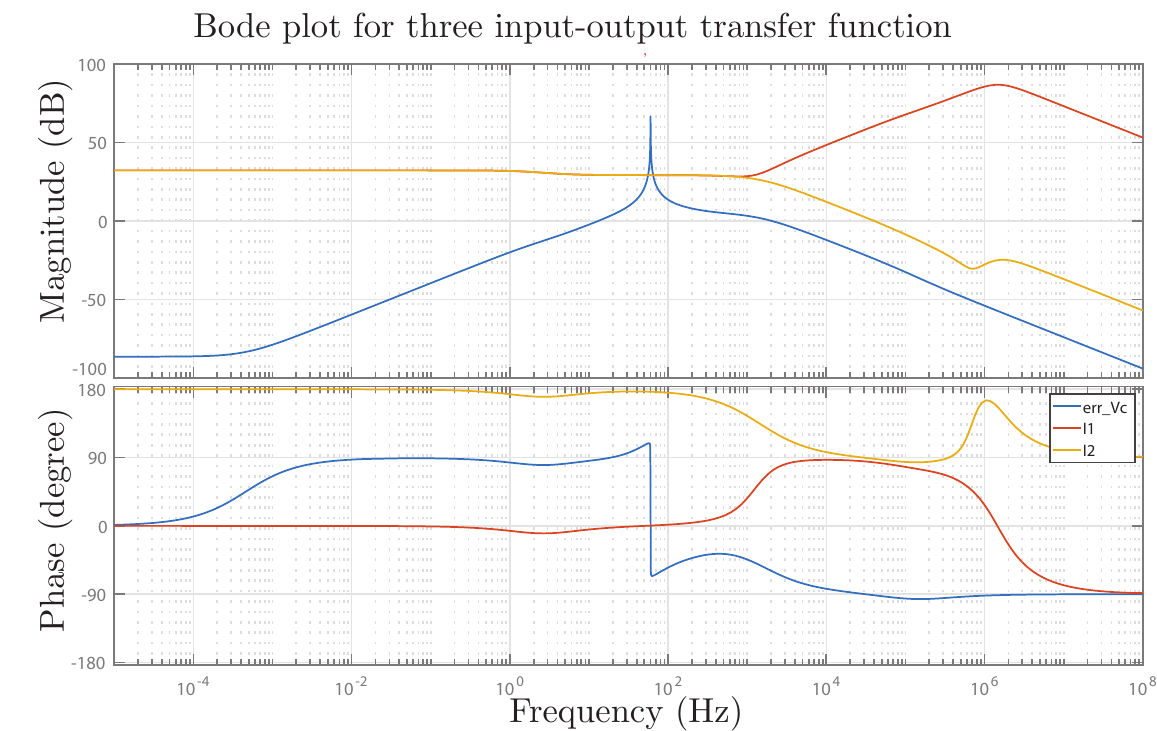}
  \end{center}
  \vspace{-10pt}
  \caption{Figure showing the Three different control input to output transfer functions} 
  \label{fig:bodeK2K3}
  \vspace{-10pt}
  \end{figure}
\subsection{Robust control formulation: Voltage control for LCL filter in Grid forming inverters}
The robust control formulation significantly differs from the one in the reference \cite{Johnson2015OptimalSF}. This is grid forming since the inverter will now define the grid frequency in the absence of any other grid or generator. The loads are modelled with constant current, $i_d$, or constant impedance, $Z_L$. We define the following equations to arrive at the state space formulation. In formulating the robust control, we will try to feed the controller the exact signal that the classical controller needs to develop the inner current-outer voltage structure. Also, we define the output of the robust controller the exact signal which the classical control generates. In essence, we give the input output signals to a black box. It will be amusing to see, how those signals, re-orient through multiple gain blocks to give the same structure as in Fig. \ref{fig:ICOVBLKdiag}.
\begin{align*}
\text{Measurement Variable} , y :\quad& v^\star-v,\quad
i,\quad
i_\text{inv}\\
\text{Regulated Variable}, z:\quad &v^\star-v,\quad
v_\text{inv}:=u+v\\
\text{Exogenous inputs}, w:\quad &v^\star,\quad i_d\\
\text{Control input}, u\quad& u:=v_\text{inv}-v
\vspace{-1cm}
\end{align*}
In light of the above description, let us take a moment to look at the A,B,C and D matrices. It follows the standard template of \cite{DnP13}, we write,
\begin{flalign*}
u:=&v_\text{inv}-v=i_\text{inv}(sL_i+R_i)\quad \\&\implies \dot i_\text{inv}=-\frac{R_i}{L_i}i_\text{inv}+\frac{1}{L_i}u\\
&v=(i+i_d)Z_L+i(sL_g+R_g)\quad \\&\implies \dot i=-\frac{R_g+Z_L}{L_g}i+\frac{1}{L_g}v-\frac{Z_L}{L_g}i_d\\
&sCv=i_\text{inv}-i\quad \implies \dot v=\frac{1}{C}i_\text{inv}-\frac{1}{C}i
\end{flalign*}
Now we arrange the following into a state space formulation as follows 
\begin{align*}
\begin{bmatrix} 
\dot i_\text{inv}\\\dot i\\\dot v
\end{bmatrix}&=\begin{bmatrix}-\frac{R_i}{L_i}&0&0\\0&-\frac{R_g+Z_L}{L_g}&\frac{1}{L_g}\\\frac{1}{C}&-\frac{1}{C}&0\end{bmatrix}\begin{bmatrix} 
 i_\text{inv}\\ i\\v
\end{bmatrix}+\begin{bmatrix}-0&0\\0&-\frac{Z_L}{L_g}\\0&0\end{bmatrix}\begin{bmatrix} 
v^\star\\i_d
\end{bmatrix}\\&
+\begin{bmatrix} 
\frac{1}{L_i}\\0\\0
\end{bmatrix}u\\
\begin{bmatrix} 
z_1\\z_2\\v_1\\v_2\\v_3
\end{bmatrix}=&\begin{bmatrix} 
0&0&-1\\0&0&1\\0&0&-1\\0&1&0\\1&0&0
\end{bmatrix}\begin{bmatrix} 
 i_\text{inv}\\ i\\v
\end{bmatrix}+\begin{bmatrix} 
1&0\\0&0\\1&0\\0&0\\0&0
\end{bmatrix}\begin{bmatrix} 
v^\star\\i_d
\end{bmatrix}+\begin{bmatrix} 
0\\1\\0\\0\\0\\0
\end{bmatrix}u
\end{align*}
\begin{figure}[b]
  \begin{center}
     \includegraphics[width=0.5\textwidth]{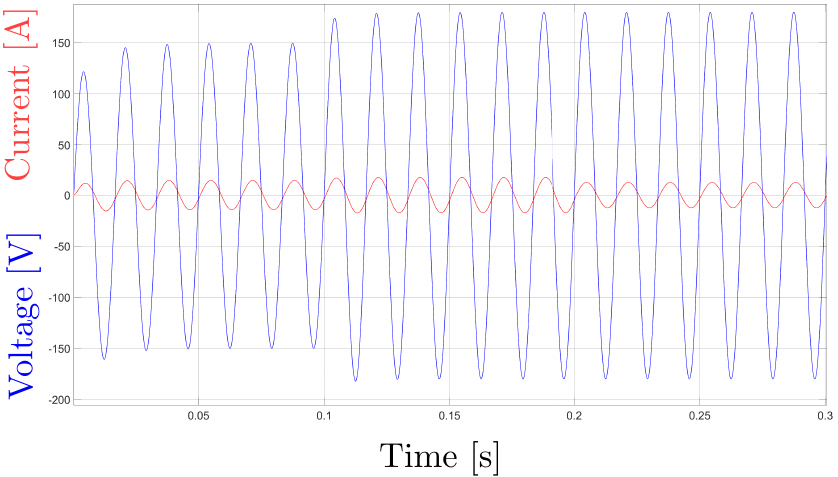}
  \end{center}
  \vspace{-10pt}
  \caption{Simulation Results}  
  \vspace{0pt}
  \label{fig:simres}
\end{figure}
Once we formulate the plant, we use MATLAB tools like \textbf{sysic} or \textbf{append} and \textbf{connect} to configure the structure of Fig. \ref{fig:generalizedhinf}. Once this is done, we specify to the \textbf{hinfsyn} that this controller has three inputs and one output and obtain the robust controller.
\subsection{Comparison to classical control structure}
To bring a one-to-one comparison with the classical ICOV structure, let us now compare from the Fig. \ref{fig:ICOVBLKdiag} and write the following equations,
\begin{align*}
    u:=v_\text{inv}-v&=K_i(i^\star_\text{inv}-i_\text{inv})\\
    &=K_i(K_ve_v+i-i_\text{inv})\\&=K_i\Big(K_v(v^\star-v)+i-i_\text{inv}\Big)\\
    &=K_iK_v(v^\star-v)+K_i(i-i_\text{inv})
\end{align*}
Comparing the equation above to the Robust control based $H_\infty$ controller we get, the following
\begin{align*}
     u:=v_\text{inv}-v&=K_1(v^\star-v)+K_2i+K_3i_\text{inv}
\end{align*}
And this is the most remarkable trait that we obtain from the results (Fig. \ref{fig:bodeK2K3}) that $K_2$ and $K_3$, are almost exactly the same in the bandwidth of concern except that $K_3$ has a phase exactly $180^\circ$ out of that of $K_2$. This means, $K_2=-K_3$. This is also immediately apparent from the equation $u=K_iK_v(v^\star-v)+K_i(i-i_\text{inv})$, that $K_i=K_2=-K_3$.\\

The following are the simulation results as shown in Fig. \ref{fig:simres}. We perform the following test on the model to check the reference tracking and disturbance rejection.
\begin{figure}[t]
  \begin{center}
     \includegraphics[width=0.5\textwidth]{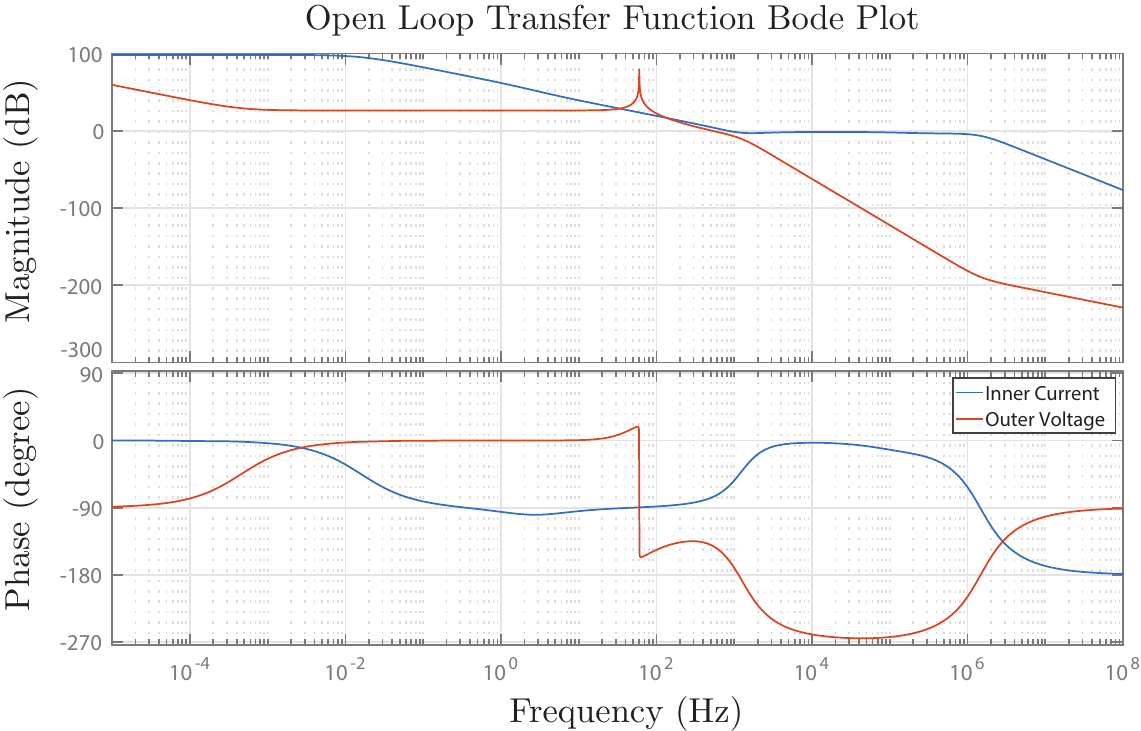}
  \end{center}
  \vspace{-20pt}
  \caption{Figure showing the different open loop plots}  
  \vspace{0pt}
  \label{fig:loop gain}
\end{figure}

\begin{figure}[t]
  \begin{center}
     \includegraphics[width=0.5\textwidth]{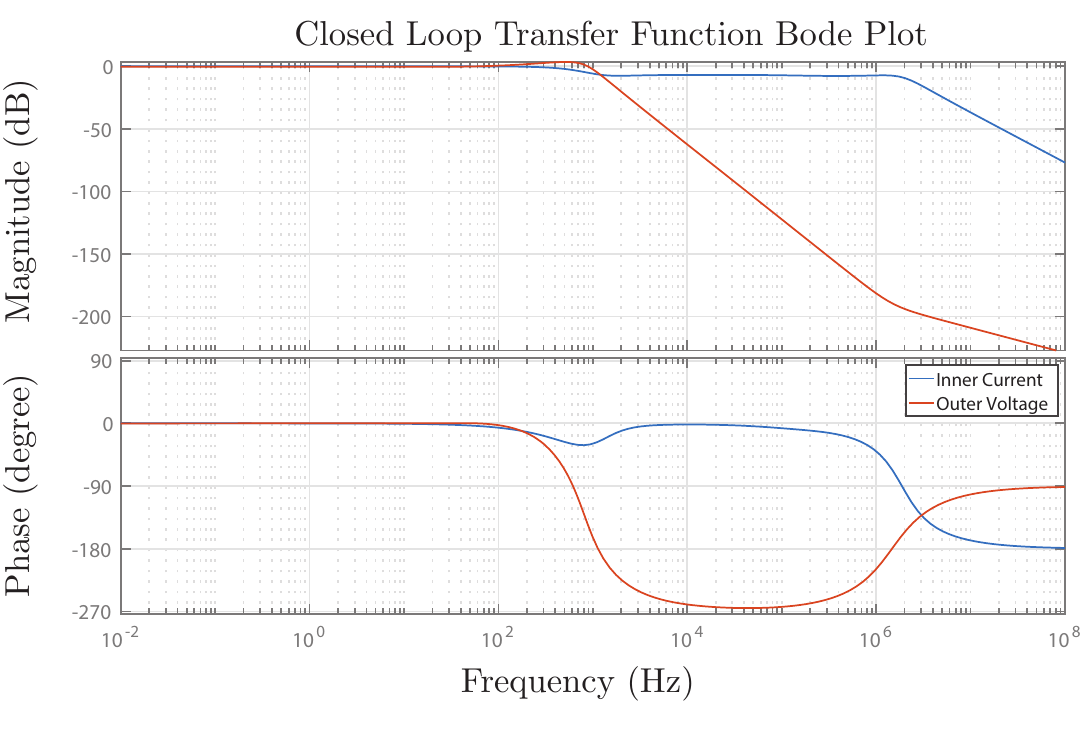}
  \end{center}
  \vspace{-10pt}
  \caption{Figure showing the closed loop plots}  
  \vspace{0pt}
  \label{fig:CLTF}
\end{figure}
\begin{figure}[b]
  \begin{center}
     \includegraphics[width=0.5\textwidth]{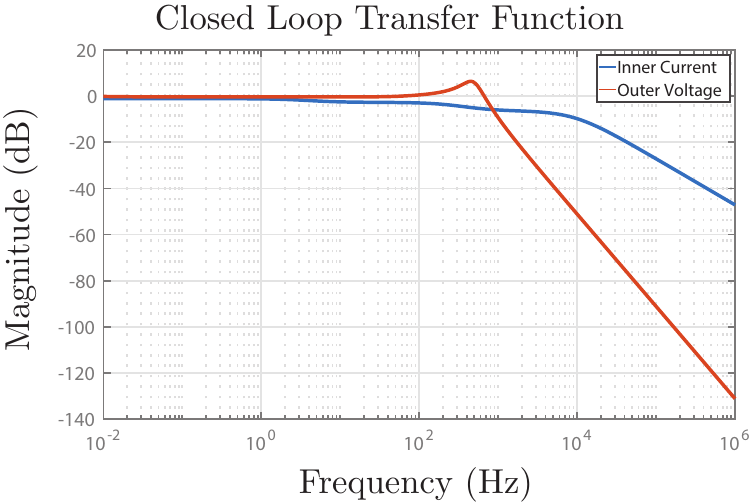}
  \end{center}
  \vspace{-10pt}
  \caption{Figure showing the closed loop plots for Case I}  
  \vspace{-10pt}
  \label{fig:caseIbode}
\end{figure}

\begin{figure}[b]
  \begin{center}
     \includegraphics[width=0.5\textwidth]{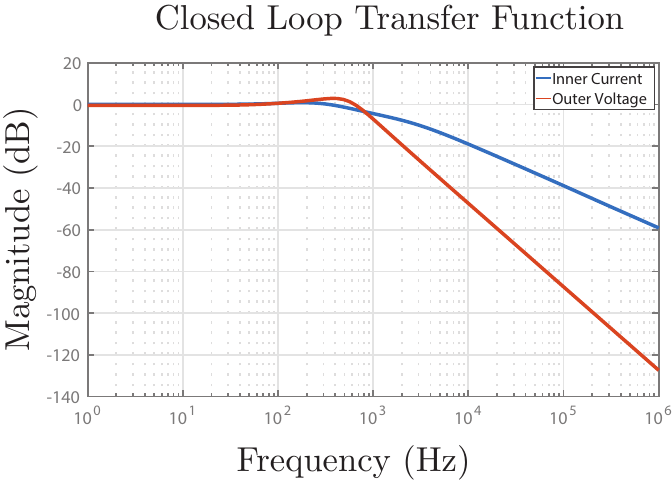}
  \end{center}
  \vspace{-10pt}
  \caption{Figure showing the closed loop plots for Case II}  
  \vspace{-10pt}
  \label{fig:case2bode}
\end{figure}
\begin{table}
\caption{Simulation: Steps of operation}
\vspace{-10pt}
\begin{center}
 \begin{tabular}{||c c c ||} 
 \hline
 Time Interval & $v^\star$ & $i_d$ \\ [0.5ex] 
 \hline\hline
 0-0.1 s&150 V &0 A \\ 
 \hline
 0.1-0.2 s&180 V& 0 A\\
 \hline
 0.2-0.5 s& 180 V& 5 A\\ [0.5ex] 
 \hline
\end{tabular}
\end{center} 
\end{table}
\vspace{0pt}
The following are the observations from Fig. \ref{fig:simres}
\begin{itemize}
    \item The reference tracking is within two cycles. This is fast enough. There is no overshoot which means that the controller has high damping. The current injection starts at 0.2s. However, there is no appreciable change in the voltage. 
    
    \item The net current flowing through the inverter grid side current now reduces by 5A since that is supplied by the disturbance current.

\end{itemize}
\subsection{Relationship between bandwidths}
This is a very important topic and probably one of the key motivation behind the project. In classical control the bandwidth relation between the inner and outer loop is always by a factor of ten since we know that the phase contribution of a pole or zero is negligible below a factor of ten. Thus when inner loop bandwidth is ten times faster, we ignore the design of inner loop in the outer loop and assume that the inner loop TF is basically unity in the frequency range of interest of the outer loop. 

This is the reason for this exercise, we try to plot the inner and the outer loop.
\begin{align*}
    \text{Inner Loop Gain} &:=\ell_i(s)=K_2(s)\frac{1}{sL_i+R_i}=K_i(s)\frac{1}{sL_i+R_i}\\
        \text{Outer Loop Gain} &:=\ell_v(s)=\frac{K_1(s)}{K_2(s)}\frac{1}{sC}=K_v(s)\frac{1}{sC}
\end{align*}
\begin{table*}[t]
\caption{Different Weighting Functions and its effect on the closed loop bandwidth ratio and feasibility of simulation}
\centering
\begin{tabular}{||c c c c c c c||}
 \hline
Case& $W_s$ & $W_u$ & $W_d$&$\omega_i/\omega_v$&Simulation Works? &$\gamma$\\ [0.5ex] 
 \hline\hline
I& $\frac{50}{s/1e3+1}\frac{s^2+2\omega_r+\omega_r^2}{s^2+0.001\,2\omega_r+\omega_r^2}\frac{s^2+2\omega_{LC}+\omega_{LC}^2}{s^2+2\,10\,\omega_{LC}+\omega_{LC}^2}$&1&1&$\approx10^{-2}$ &NO (Fig. \ref{fig:caseIbode},\ref{fig:simcase2})& 6.2828\\ 
 \hline
II&$\frac{50}{s/1e3+1}\frac{s^2+2\omega_r+\omega_r^2}{s^2+0.001\,2\omega_r+\omega_r^2}\frac{s^2+2\omega_{LC}+\omega_{LC}^2}{s^2+2\,10\,\omega_{LC}+\omega_{LC}^2}$&$0.01\,\frac{s/100+1}{s/10^6 +1}$&1& $\approx1$& NO (Fig. \ref{fig:case2bode})&4.9437\\
 \hline
 III& $\frac{50}{s/1e3+1}\frac{s^2+2\omega_r+\omega_r^2}{s^2+0.001\,2\omega_r+\omega_r^2}\frac{s^2+2\omega_{LC}+\omega_{LC}^2}{s^2+2\,10\,\omega_{LC}+\omega_{LC}^2}$ &1& $\frac{1}{s/3147+1}$ & $\approx10^3$& YES&4.566\\ 
 \hline
IV&$\frac{50}{s/1e3+1}\frac{s^2+2\omega_r+\omega_r^2}{s^2+0.001\,2\omega_r+\omega_r^2}\frac{s^2+2\omega_{LC}+\omega_{LC}^2}{s^2+2\,10\,\omega_{LC}+\omega_{LC}^2}$&$0.01\,\frac{s/100+1}{s/10^6 +1}$& $\frac{1}{s/3147+1}$&  $\approx10^3$ & YES (Fig. \ref{fig:simres},\ref{fig:CLTF})&2.395\\ [0.5ex] 
 \hline
\end{tabular}
\end{table*}
Fig. \ref{fig:loop gain} shows the loop gains of the two transfer functions. Whereas, Fig. \ref{fig:CLTF} shows the closed loop TF which is basically $\ell_i/(1+\ell_i)$ and $\ell_v/(1+\ell_v)$ respectively. We define the bandwidths to be $\omega_i$ and $\omega_v$ respectively.

Fig. \ref{fig:CLTF} shows that the bandwidth of these two loops is separated by a factor greater than the rule of thumb or 10. 

The following questions are important.
\begin{itemize}
    \item [$\square$] How does the bandwidth of the generated controllers relate to the weighting functions.
        \item [$\square$] How does the bandwidth of the generated controllers relate to the value of $\gamma$.
            \item [$\square$] How does the controller actually does the trick to generate the ICOV structure. What is the ARE/LMI or weighting function that is forcing to do this.
\end{itemize}

\begin{figure}[t]
  \begin{center}
     \includegraphics[width=0.5\textwidth]{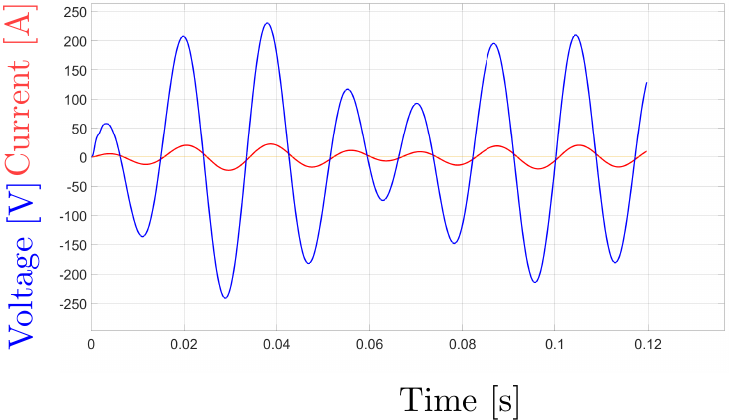}
  \end{center}
  \vspace{-10pt}
  \caption{Figure showing the response of the plant to the same kind of reference as we had subjected in Table I}  
  \vspace{-10pt}
  \label{fig:simcase2}
\end{figure}
We have not been able to come to any direct conclusion or a good understanding of these questions, but the following points might help.

\subsection{Observations from the Table II}
We can make some observations from the different weighting functions. First and foremost is that we do need the three weighting functions to be specified. If we do not specify them, we get unstable controller. It also, very intuitively gives us the conclusion that the plant is stabilized only when the inner control loop is faster than the outer one. We knew this. The only thing we wanted to find was, how much faster?

These are some interesting conclusions

\begin{itemize}
    \item We are often accustomed to having resonant controller in both the inner and the outer controller. Fig. \ref{fig:loop gain} shows that the inner controller can as well only be a proportional controller. This technique has well been used in literature perhaps in an ad-hoc way. The general norm is, if the outer loop takes care of steady state error by integrator or resonant controller, the inner loop might not as well have that. We come to the same point, only through robust control techniques.
    
    \item We observe that the weighting function $W_s$ very strongly ties to the error in voltage reference. We can thus claim that the zero crossing of the weighting function influences the bandwidth of the closed loop bandwidth of the outer voltage controller ($\ell_v$).
    \item The question then remains, is to see that what determines the current loop bandwidth. We observe that the weighting function $W_d$ related to the disturbance in current. Thus it might be true that changing the weighting function, $W_d$ might change the variable current controller bandwidth. The present $W_d$ is, $\frac{1}{s/3147+1}$. This signifies that all the disturbances upto the frequency 10 times higher than maximum disturbance will be tried to be rejected. Thereafter, to reduce sensitivity to noise, we need to roll-off $W_d$. The minimum we can reduce this is to slightly higher than $2\,\pi\,60=377 \text{rad/s}$. This $W_d$ still gives a ratio of $\omega_i/\omega_v$ to be very high, typically in order of $10^4$. Only when we reduce the pole of $W_d$ significantly low, to be around the following $W_d$, $\frac{1}{s/10^{-4}+1}$ do we get to a point where the ratio of $\omega_i/\omega_v$ is much lower, around $10^{-2}$. Now when we simulate the plant with this controller, we get an oscillatory and unstable system.
    \end{itemize}

   \section{Conclusion and Scope for Improvement} \label{sec: conclusion}
   
 In this paper we have demonstrated the use of the weighting function in synthesizing controllers for the power converters for various filter and signals (ac and dc). We have discussed the design and selection of these weighting functions and shown their performance over classical controller with cogent simulations. Finally we took up the interesting problem to show how a complicated intuition-based classical controller structure comes from the rigorous mathematics of the robust controllers. Through various test cases, we show that the bandwidth ratio can only be pushed to around $10^3$, much higher than what was already known from the classical control. 
 
 The main drawback of this current work is not having a clear understanding of the clear ties between the weighting functions and the controller synthesis when there are multiple weighting functions and we are dealing with a MIMO system. If we had a better mathematical model, we could have obtained some valuable information from the $\gamma$ and commented upon a theoretical minima on the $\omega_i/\omega_v$ ratio.

{
\footnotesize
\bibliographystyle{ieeetr}
\addcontentsline{toc}{section}{\refname}
\bibliography{bibliography}
}
\end{document}